\documentclass[11pt,oneside]{article}
\usepackage{amsmath}
\usepackage{amsfonts}
\usepackage{amssymb}
\usepackage{latexsym}
\usepackage{graphicx}

 \advance\textheight by \topskip
\large\normalsize
\newcommand{\be}{\begin{equation}}
\newcommand{\ee}{\end{equation}}
\newcommand{\ba}{\begin{array}}
\newcommand{\ea}{\end{array}}

\newcommand{\pa}{\partial}

\begin{document}

\centerline{hep-th/0107141 \hfill CERN--TH/2001--189}
\vspace{5mm}
\vspace{0.5cm}
\centerline{\LARGE\bf Holography and the Electroweak Phase Transition}

\medskip\bigskip
\centerline{\large\bf Paolo Creminelli, Alberto Nicolis}
\centerline{\em Scuola Normale Superiore and INFN, Sezione di Pisa, Italy}
\vspace{3mm}
\centerline{\large\bf Riccardo Rattazzi\footnote{On leave from INFN, Pisa, Italy.}}
\vspace{0mm}
\centerline{\em Theory Division, CERN, Geneva 23, CH-1211, Switzerland}
\vspace{1cm}
\centerline{\large\bf Abstract}
\begin{quote}\indent 
We study through holography the compact Randall-Sundrum (RS) model at finite 
temperature. In the presence of radius stabilization,  the system is 
described at low enough temperature by the RS solution. At  high 
temperature it is described by the AdS-Schwarzschild solution with an event horizon replacing the TeV brane.  
We calculate the transition temperature $T_{c}$ between the two phases and we find it to be  somewhat smaller 
than the TeV scale. 
Assuming that the Universe starts out at $T\gg T_{c}$ and cools down by 
expansion, we study the  rate of the transition to the RS phase. 
We find that the transition is very slow so that an inflationary phase at the weak scale begins.  
The subsequent evolution depends on the stabilization mechanism: in the simplest Goldberger-Wise case 
inflation goes on forever unless tight bounds are satisfied by the model parameters; 
in slightly less-minimal cases these bounds may be relaxed.   

\end{quote}

\section{Introduction}
In recent years particle physics has been shaken by the realization that
new compact space dimensions could be so large to be accessible in
collider experiments \cite{anto,add,aadd,rs1}. This realization allows 
for a new viewpoint
on the open problems of particle physics and most notably on the hierarchy problem.
In particular in the scenario proposed by Arkani-Hamed, Dimopoulos and Dvali (ADD) \cite{add} the 
fundamental scale of quantum gravity is of the order of the weak scale. A crucial feature
of this proposal is that the Standard Model degrees of freedom should be localized on a 
3-dimensional defect, a {\it brane}, so that they cannot access directly the new dimensions. 
Then the weakness of 4D
gravity originates from the large radius $R$ of compactification: solving the hierarchy 
problem would now consist in explaining why $1/R$ is so much smaller than the weak scale. 
Randall and Sundrum (RS) \cite{rs1} have instead proposed a warped compactification with one
extra dimension. The full space-time is a slice of AdS$_5$ bordered by two branes. The interest 
of this model is in that it can elegantly explain the hierarchy as due to gravitational
red-shift of energy scales. Also in this case  the hierarchy depends on
the size of the extra dimension, but Goldberger and Wise (GW) \cite{Goldberger:1999uk} have 
shown a simple mechanism that can naturally generate the right order of magnitude.

The presence of new space dimensions dramatically changes the early evolution of our
Universe. When the temperature is higher than $1/R$, Kaluza-Klein (KK) particles
can be produced with important cosmological consequences, not always pleasant.  
Indeed in the ADD scenario
the KK gravitons produced at high temperature tend to over-close the Universe and, upon decaying
on the brane, to distort the diffuse photon spectrum. This places a rather strong bound of
about 1-100 MeV on the maximal temperature at which the Universe was ever 
heated \cite{add}. With such a low maximal 
temperature  the standard Big Bang Nucleosynthesis barely fits into the
picture. It is  regarded as rather unnatural that standard 
cosmology should start right at the time of nucleosynthesis.
In other models, like RS or ref.~\cite{anto}, the maximal temperature at which the Universe is surely
normal is about 1 TeV. This scale represents the separation of KK levels, and anyway it is about
the scale where the interactions of these extra-dimensional theories go out of perturbative control
and the effective field theory description breaks down.
$T_{\rm max} \sim 1$ TeV does not pose any direct threat to the standard Big Bang, though it severely 
constrains the building of models of inflation or baryogenesis.
Therefore, provided it existed, an alternative description valid in the regime $T\gg 1$ TeV
would undoubtedly be worth investigating. The purpose of this paper is to do so in
the RS model, for which such a high energy description is offered by the AdS/CFT 
correspondence \cite{maldacena}. 

According to the AdS/CFT interpretation \cite{maldacena2} the RS
model represents a quasi-conformal, strongly coupled 4D-field theory 
coupled to 4D-gravity. In this view, the KK resonances,  the graviton zero mode excluded, and the 
Standard Model particles are just bound states of this purely four dimensional theory
\cite{Arkani-Hamed:2000ds}. This viewpoint clarifies the puzzles raised by the RS model. 
For instance, the CFT picture
explains in a simple way why an observer at the SM boundary sees gravity becoming strong at 1 TeV,
while an observer at the other boundary only experiences quantum gravity at energies of
order $10^{19}$ GeV: the observer at the SM boundary is part of the CFT, the strong coupling 
he experiences at 1 TeV is analogous to what pions go through at 1 GeV and is not truly due
to gravity becoming strong. The CFT interpretation of the RS model is valid pretty much
for the same reasons that gravity on full AdS$_{d+1}$ space describes a $d$-dimensional CFT without 
gravity \cite{Witten:1998qj}.  Several qualitative and quantitative checks of the consistency of
the holographic interpretation have been given \cite{Gubser:2001vj,verlinde1,Arkani-Hamed:2000ds,
Rattazzi:2001hs}. In absence of a stabilization mechanism, the RS model is a CFT where dilation
invariance is spontaneously broken, the radion $\mu$ being the corresponding Goldstone boson (dilaton).
The GW stabilization mechanism is holographically dual to turning on a quasi marginal
deformation of the CFT which generates a small weak scale by dimensional transmutation.
This is analogous to what happens in technicolor models.

In this paper we  study the finite temperature behavior of the RS model from the holographic
perspective. In the gravity picture now time is Euclidean and compactified on a circle, while the 5D-gravitational
action is interpreted as the free energy of the corresponding 4D theory.
We  argue, as suggested in ref.~\cite{Arkani-Hamed:2000ds}, that at high enough temperature the model is in a phase described 
on the gravity side by AdS$_5$-Schwarzschild (AdS-S), with the TeV brane replaced by the black-hole
horizon. This is in close analogy to what happens in the standard
AdS/CFT correspondence at finite $T$ \cite{Witten:1998zw}. Indeed in
absence of a stabilization of the 5th dimension the dual 4D theory is conformal (though spontaneously broken) and there is no
distinction between high and low temperature: at the gravity level it is 
described by AdS-S at any finite temperature. This picture is based on 
some test calculation and on entropic considerations.
For example we calculate around the RS classical solution with the TeV brane the leading thermal correction to the radion 
potential in the regime $\mu\gg T$. In the absence of stabilization, we find 
that $\mu$ is pushed towards $\mu\sim T$ where 
5D quantum gravity effects becomes strong and where we expect the AdS-S  black-hole  to be formed. 
On the other hand when the radion is stabilized, at low enough temperature, the usual RS solution with the TeV brane and
no horizon is  perturbatively stable. The action of this solution (free energy) essentially equals the GW potential
at its minimum. Comparing the action of the two solutions we determine the critical temperature $T_c$ at which 
a first order phase transition takes place: for $T>T_c$ the theory is in the hot conformal phase (AdS-S) while for $T<T_c$
the theory is in the SM phase (RS model with TeV brane).
 We find that $T_c$ is somewhat lower that the weak scale 
 unless the GW scalar background causes a big distortion of the RS metric.
As we discuss, such a low critical temperature causes the phase
transition to the SM phase as the Universe cools down from a primordial hot CFT phase to be very slow. 
In the simplest implementation of the GW mechanism \cite{Goldberger:1999uk}, unless
the 5D Planck scale $M$, the AdS curvature $1/L$ and the GW scalar  
profile $\phi^{{2/3}}$ are comparable, thus spoiling the small curvature expansion,
the Universe inflates in the CFT forever as in the old inflation scenario.
In this case, even if a high temperature description of the RS model is available, to obtain a 
viable cosmology we are forced to suppose that the Big Bang temperature does not exceed the weak scale, 
similarly to the models with flat TeV scale extra dimensions. This behaviour can be traced back to the presence,
in the minimal GW scenario, of a secondary minimum at $\mu \sim 0$. By modifying the stabilization
mechanism, we can obtain radion potentials without a secondary minimum: in this case we have 
a cosmological scenario in which, after a long period of inflation, the phase transition finally happens. 
Even if the reheating temperature remains bounded by the TeV scale, we may have a 
predictive cosmological evolution starting from temperatures as high as the Planck scale.  

This paper is organized as follows. In section 2 we give our basic notation and conventions and present an illustrative
computation of the radion effective potential at low temperature. In sections 3 to 6 we discuss the two gravitational
solutions that are relevant at finite $T$, including the effects of
a GW scalar field. In sections 7 and 8 we  study the dynamics of the phase transition.
We give a rough estimate of the rate of bubble nucleation and compare it to the expansion rate of the Universe.
In section 9 we discuss our results. In the appendix we present instead a solution where a ``fake'' horizon
is hidden behind the TeV brane. This solution, though it  corresponds to zero temperature (no horizon), it gives
the same FRW evolution of a radiation dominated Universe: this result is easily understood as due to
the conformally coupled radion.

\section{The RS model at finite temperature     \label{sec:dest}}
We first briefly
recall the definition of the model and establish our notation. The fifth dimension is
compactified on an
$S_1/Z_2$ orbifold. We parameterize the single covering of $S_1/Z_2$ with a coordinate $z \in [z_0,z_1]$.
The 4D subspaces at the $z_0$ and $z_1$ boundaries are respectively the Planck brane and the TeV brane. The pure
gravity part of the action is
\be
\label{eq:5Daction}
S = 2 M^3 \int d^4 xdz \:\left[\sqrt{-g}( R + 12
k^2)-12k\sqrt{-g_0}\;\delta(z-z_0)+12k\sqrt{-g_1}\;\delta(z-z_1)\right ] \;, 
\end{equation}
where $M$ is the 5D Planck mass, $k=1/L$ is the AdS curvature, $g_{0}$ and $g_1$ are the induced metric at the
two boundaries and the associated terms can be interpreted as the
tension of the corresponding branes. Then the metric
\be
\label{eq:rsmetric}
ds^2 = e^{-2kz}\eta_{\mu\nu} dx^\mu dx^\nu + dz^2
\end{equation}
solves Einstein's equations over the full space, including the jump discontinuities at the 
boundaries\footnote{For more general solutions see Appendix \ref{radiondriven}.}. Due to
the warp factor in the metric,  a 4D  theory localized at $z_1$  experiences a redshift factor
$e^{-k(z_1-z_0)}$ in its energy scales with respect to a 4D theory at $z_0$. Conversely, as the 4D graviton
mode is localized near the Planck brane,
the four dimensional Planck scale is not redshifted
\be
\label{eq:4dplanckscale}
M_4^2=M^3L(e^{-2kz_0}-e^{-2kz_1})\;.
\end{equation}
So if the Standard Model is
localized at $z_1$ the electroweak hierarchy can be explained for 
$(z_0-z_1)/L\sim 30$ keeping $M\sim
1/L$ \cite{rs1}. The size of the fifth dimension is however a modulus, corresponding to a massless field: the
radion
$\mu$. To truly explain the hierarchy this degeneracy must be lifted. The simplest solution to this problem
is the GW mechanism that we will discuss later on.

The AdS/CFT correspondence offers a very useful 4D interpretation of the RS model. By the AdS/CFT
correspondence, gravity (or better string theory) on full
AdS$_{p+1}$ is dual to a $p$-dimensional conformal field theory without gravity. The correspondence is a duality
since in the limit  $ML\gg 1$ (and $g_{\rm string} \ll 1$) where classical gravity is a good description on the AdS
side, the corresponding CFT is strongly coupled (large number of states $N^2$ for large and fixed 't~Hooft
coupling). One of the essential aspects of this duality is that the isometries of AdS$_{p+1}$ form the group $SO(p-1,2)$,
which coincides with the conformal group in $p$-dimensional spacetime. For example the isometry $z\to z+ L\ln \lambda$,
$x\to \lambda x$ corresponds to a dilation in the $p$ dimensional CFT. In particular notice that a shift to a larger $z$
corresponds to going to larger length scales in the $p$-dimensional theory. Therefore from the holographic viewpoint
going  to smaller (larger) $z$ is equivalent to RG evolving towards the UV (IR) in the lower dimensional theory.
In this respect the RS model looks like a 4D theory well described by a CFT at intermediate energies, but whose 
UV and IR behavior deviates from a pure CFT  because of the presence of respectively the Planck and TeV brane. Now, by
taking the limit  $z_0\to -\infty$, the Planck brane
is (re)moved to the boundary of AdS. In this limit  $M_4\to \infty$ and 4D gravity decouples.
However the spectrum of KK gravitons is not much affected, their mass splittings being  set by the radion vacuum
expectation value (VEV)  $\langle \mu\rangle=ke^{-kz_1}$. Unlike in the ADD scenario, there remains no
4D long distance force mediated by spin 2 particles, as if there was really no gravity. This is consistent with a
purely 4D interpretation of the model. The resulting model with just the TeV brane  is directly interpreted as a 4D theory
where conformal invariance is spontaneously broken by the VEV of the radion. Notice for instance that $\langle
\mu\rangle=ke^{-kz_1}$, which fixes the mass scale, transforms with conformal weight  1 under the AdS dilation isometry
mentioned above: a clear sign of spontaneous breaking. Finally, bringing in from infinity  the Planck brane with its
localized graviton zero mode corresponds to gauging 4D gravity in this CFT.

Let us now study the thermal properties of the model.  While at zero
temperature $\mu=ke^{-kz_1}$ is a flat direction, at $T\not = 0$ we expect  a potential to be generated.
For simplicity let us decouple 4D gravity by moving the Planck brane towards 
$z=-\infty$ and consider a toy model where the TeV brane is empty. To further simplify let us also focus on
$T \ll \mu$. In this limit the KK are too heavy to be thermally excited and the radion
is the only relevant degree of freedom. Virtual KK exchange is however the source of radion 
interaction. In order to calculate the radion effective potential we must first integrate out the KK modes
to get an effective Lagrangian. The leading effects arise from tree level KK exchange. To study 
fluctuations around the RS metric it is 
convenient to choose coordinates where the TeV brane is not bent and located 
at $z_{1}$ and the metric given by
\be
\label{eq:rubypert}
ds^2 = e^{-2kz-2f(x)e^{2kz}}(\eta_{\mu\nu} + h_{\mu\nu}(x,z)) dx^\mu dx^\nu + (1+2f(x)e^{2kz})^2 dz^2 \;.
\end{equation}
In this particular parameterization there is no
kinetic mixing between the scalar perturbation $f(x)$ (radion) and the spin 2 fluctuations $h_{\mu\nu}$ 
\cite{rubakov}. The kinetic term for the radion turns out to be 
\cite{Csaki:2000mp,Goldberger:2000un}
\be
{\cal{L}}_{\rm kin} = -12 (M L)^3 e^{2kz_{1}Ç} k^2 (\partial_\mu f)^2        \; .
\end{equation}
Notice that in the standard parameterization the radion $\mu$ is 
given by the warp factor at the TeV brane: $\mu=k \;{\rm exp}(-k z_{1}- 
fe^{2kz_{1}})$. In what follows we will treat $f$ as the quantum fluctuation over the  background radion VEV $\langle
\mu\rangle=ke^{-kz_1}$. At low energy (low temperature) the leading 
radion interaction term involves 4 derivatives, and it arises by KK mode exchange in the 
static limit (neglecting four momentum in their propagator). By substituting eq.~(\ref{eq:rubypert}) into 
eq.~(\ref{eq:5Daction}), the relevant action 
terms are
\be
\label{eq:relevant}
{\cal{L}}(x) = 2M^3 \int dz \left\{e^{2kz} h_{\mu\nu} T^{\mu\nu} + e^{-4kz} \frac{1}{4} 
(-\pa_z h^{\mu\nu} \pa_z h_{\mu\nu} + \pa_z h^{\mu}_{\mu} \pa_z h_{\nu}^{\nu})\right\} 
\end{equation}
\be
\label{eq:T}
T^{\mu\nu}(x) \equiv 2\pa^\mu f(x) \pa^\nu f(x) - 4f(x) \pa^\mu\pa^\nu f(x) 
+ \eta^{\mu\nu} \left(\pa^\rho f(x) \pa_\rho f(x) + 4 f(x) \pa^\rho\pa_\rho f(x)\right).   
\end{equation}
The first term in ${\cal{L}}(x)$ describes the interaction between two radions and a graviton, while the second one
is the five dimensional contribution to the graviton kinetic  term. As expected, $T^{\mu\nu}$ is proportional 
to the radion stress-energy tensor. The coupling of $f$ to 4D gravity is non minimal because of the 
term
$- \sqrt{-g} \xi R f^2/2$ with $\xi =1/3$ \footnote{Notice that for the field $\mu=k\;{\rm exp}(-k z_{1}- 
fe^{2kz_{1}})$ the non-minimality parameter is
\cite{Goldberger:2000un} $\xi = 1/6$, as required by conformal invariance \cite{Rattazzi:2001hs} 
(see also Appendix \ref{radiondriven}). The 
point is that $\mu$ transforms with  weight 1 under Weyl rescaling, while $f$ transforms non linearly.}.

By integrating out the gravitons at tree level we get 
\be
\label{eq:intout}
{\cal{L}}_{\rm eff} (x) = 2 M^3 \int dz dz' e^{2kz} e^{2kz'} G(z,z') \left[-T^{\mu\nu} T_{\mu\nu}
+ \frac{1}{3} T_\mu^\mu T_\nu^\nu \right],
\end{equation}
where $G(z,z')$ is the scalar Green function over the RS metric satisfying $(\pa_z e^{-4kz} \pa_z) G(z,z') = \delta(z-z')$
with Neumann boundary conditions at $z=z_{1} $. $G$ is given 
by\footnote{The normalization is the one for the double covering of 
the $(-\infty,z_{1})$ region. }
\begin{eqnarray}
G_<(z,z') & = & - \frac{1}{8k} e^{4kz} \qquad z > z' \\
G_>(z,z') & = & - \frac{1}{8k} e^{4kz'} \qquad \!\!z < z' \;.
\end{eqnarray}
Finally, integrating over $z$ and $z'$ we arrive at
\be
\label{eq:effectiveL}
{\cal{L}}_{\rm eff} (x) = (ML)^3 \frac{e^{8kr}}{24} \left[T^{\mu\nu} T_{\mu\nu}
- \frac{1}{3} T_\mu^\mu T_\nu^\nu \right] \;,
\end{equation}
which describes a 4-derivative interaction among four radions. Trilinear 
couplings are absent in our  parameterization as there is no radion-graviton kinetic mixing. 
The leading correction to the radion  potential is the thermal average of $-{\cal{L}}_{\rm eff}$. The only
relevant term will be proportional to $\langle f \pa^\mu \pa^\nu f\rangle \langle f \pa_\mu \pa_\nu f \rangle$: 
all the other averages 
are not $T$ dependent being proportional to the equations of motion. In terms of the canonically 
normalized  radion $\tilde f$ we have
\be
\label{eq:Vterm}
V_T (\mu) = - \frac{1}{(M L)^3} \frac{1}{288} \frac{1}{\mu^4} \langle \tilde f \pa^\mu \pa^\nu 
\tilde f \rangle \langle \tilde f \pa_\mu \pa_\nu \tilde f \rangle.
\end{equation} 
Notice that  $\langle \tilde  f \pa^\mu \pa^\nu \tilde f \rangle$ is, up to $T$ independent pieces,
the stress energy tensor of a free scalar. In the end the result is 
\be
\label{eq:termpot}
V_T (\mu) = - \frac{1}{(M L)^3} \frac{\pi^4}{194400} \frac{T^8}{\mu^4}.
\end{equation}
Eq.~(\ref{eq:termpot}) tells us that for $\mu\gg T $ the TeV brane is destabilized and pushed by thermal effects 
towards the AdS horizon\footnote{This instability is somewhat related to the Jeans instability of flat space 
at finite temperature. Also the quantum instability we will describe in the following may be compared 
to the
quantum production of black holes from flat space. These gravitational instabilities at finite temperature
are analyzed in \cite{Gross:1982cv}.}. 
This result could have been anticipated. The $T^{8}/\mu^{4}$ behavior 
is both due to  conformal 
invariance (or, equivalently, dimensional analysis) and to the 4-derivative character of the interaction. 
The  minus sign is due to the attractive nature of forces mediated by spin 2 particles.
As $\mu$ is pushed below $T$,  KK-gravitons start being thermally produced. In the regime 
$T>\mu>T/(M L)$ the free energy  goes roughly like $-(T/\mu) T^{4} \sim T^{5}$, where $T/\mu$ is the number of 
thermally
excited gravitons. When $M L \mu=\tilde M_{5} \sim T$,  perturbation theory breaks 
down: for such value of $\mu$ the temperature is Planckian for a TeV brane observer. Equivalently,
the length $e^{-kz_1}/T$ of the time cycle in Euclidean space  becomes Planckian.  To summarize: 
the radion thermal potential has the form $V=T^{4} g(T/\mu)$, it is monotonically 
decreasing with $\mu$ in the regime  $\mu > T/(M L)$, and therefore  thermal effects drive the  RS model 
into the Planckian regime. We conclude that around the RS solution thermal equilibrium is either impossible
or described by messy Planckian physics. 

Fortunately gravity itself provides an elegant way out of this situation. As pointed out by Hawking and Page 
\cite{Hawking:1983dh},  
the canonical ensemble of AdS space is described by the AdS-Schwarzschild (AdS-S) solution. In this case Hawking
radiation from the  black-hole horizon allows a hot bulk to be in equilibrium. It is then natural to assume
that AdS-S will also describe the thermal phase of the RS  model with the TeV brane  replaced by the 
horizon. The qualitative picture is then the following: thermal radiation falls with the TeV brane towards $z=\infty$
until the energy density (see discussion above) is so large that everything collapses
to form a black hole \cite{Hebecker:2001nv}. 
This way we do not have to worry about Planckian physics as the Schwarzschild
horizon acts as a censor.

We now discuss this in more detail also considering the relevant case of a stabilized radius.



\section{Two gravity solutions}
Let us restart and consider the RS model at finite temperature from the viewpoint of  the AdS/CFT correspondence.
AdS/CFT  relates a conformal theory defined on a $d$-dimensional manifold ${\cal{M}}$ to gravity (string theory) 
defined on the product of a ($d$+1)-dimensional space ${\cal{X}}$ and a compact manifold ${\cal{W}}$. ${\cal{X}}$ 
is an Einstein manifold
with  negative cosmological constant and its boundary must coincide with ${\cal{M}}$.
The precise relationship between the two theories is given through the field-operator correspondence 
\cite{Witten:1998qj}
\be     \label{eq:correspondence}
\left\langle \exp \int_{\cal{M}} \phi_0 \cal{O} \right\rangle _{\rm CFT} = Z_S (\phi_0)     \; ,
\end{equation} 
where the field $\phi$ on the gravity side corresponds to the operator $\cal{O}$ in the CFT and it reduces
to $\phi_0$ on the boundary of ${\cal{X}}$. The string partition function $Z_S$ is replaced at low energies by the
corresponding (super)gravity quantity.
The crucial remark now \cite{Witten:1998qj,Witten:1998zw} is that when several spaces ${\cal{X}}_i$ have 
the same boundary ${\cal{M}}$ the partition 
function $Z_S$ must be replaced by a sum over all these possibilities. 
This point is used in \cite{Witten:1998qj,Witten:1998zw} to give the dual gravitational description of the
deconfining phase transition of a CFT defined on a sphere; two different gravitational descriptions are relevant: 
one of them gives the dominant contribution to the partition function at high temperature, while the other 
dominates 
at low temperature. The transition between these two gravitational regimes was described by Hawking and Page 
\cite{Hawking:1983dh} in the context of quantum gravity and is now seen as dual to the CFT phase transition.  

Similar considerations turn out to be relevant in our problem. At finite temperature the manifold ${\cal{M}}$ 
is in our case  ${\mathbb R}^3 \times S^1$ and two different ``bulk'' spaces must be considered: 
one is the RS solution with 
TeV brane  and time compactified (we keep the Planck brane at infinity), while the other is the 
AdS-Schwarzschild solution (AdS-S). These are the only solutions describing states of thermodynamical equilibrium:
there are more generic non-thermal solutions as we will see in the Appendix. The AdS-S Euclidean metric is  
\be
\label{eq:AdSS}
d s^2 = \left(\frac{\rho^2}{L^2} - \frac{\rho_h^4/L^2}{\rho^2}\right) d t^2 + 
\frac{d \rho^2}{\frac{\rho^2}{L^2} - \frac{\rho_h^4/L^2}{\rho^2}} + \frac{\rho^2}{L^2} \sum_i d x_i^2   
\end{equation}
for $\rho_h \leq \rho < \infty$. For $\rho_h = 0$ this reduces to the pure AdS metric 
\be
\label{eq:AdS}
d s^2 = \frac{\rho^2}{L^2} \left(d t^2  + \sum_i d x_i^2 \right) + \frac{L^2}{\rho^2} d \rho^2 \;,  
\end{equation}
equivalent to (\ref{eq:rsmetric}) with $\rho = L \exp (-z/L)$. 

This solution describes a black hole in AdS space
with event horizon situated at $\rho = \rho_h$
(\footnote{Notice that putting a 3+1 brane
at the Schwarzschild horizon is equivalent to putting none. This is simply 
because, being infinitely redshifted, it would not affect the dynamics. This
is very clear in euclidian space, where the 4-brane would be shrunk to
a 3-dimensional plane.}).
The metric (\ref{eq:AdSS}) is solution 
of the Einstein equations only for a specific value of the time periodicity $\beta$,
\be
\label{eq:Th}
\beta^{-1} =  \frac{\rho_h}{\pi L^2}\equiv T_h   \; ;
\end{equation}
the temperature $T=\beta^{-1}$ must be equal to the Hawking temperature of the black hole. For $T \neq T_h$ a 
conical singularity arises at the horizon.

The real time 4D interpretation of this solution is quite simple \cite{Gubser:2001vj}: 
the Universe is filled with a thermal CFT state with  temperature $T_h$. The energy and entropy of the heat bath 
are described by the corresponding quantity of the black hole. Moreover 
if the space is ended
at some $\rho_0<\infty $ by the Planck brane, then $\rho_0$ has  to depend 
on the proper time on the brane in order to satisfy the Israel junction conditions. 
This way the induced metric on the Planck brane varies according to a radiation dominated Universe. 

A phase transition will occur between the two regimes: there is no real 
time evolution between the two gravitational solutions as somehow hinted in
\cite{Arkani-Hamed:2000ds}. Note that our problem is similar to the transition between AdS and AdS-S
described by Witten \cite{Witten:1998zw}: there the explicit breaking of conformal invariance is given by
the radius of the 3-sphere on which the CFT lives, while in our case it is given by the GW mechanism.
Being interested in understanding which of the above bulk solutions dominates the partition function at a generic 
temperature $T$, in the following sections we compute the free energy of both solutions
including the effects of a GW scalar field.

\section{The free energy of the black hole solution}
We proceed to compute the free energy of the AdS-S solution (\ref{eq:AdSS}) at temperature $T$. As the following 
results will be useful when studying the dynamics of the phase transition, we allow the horizon to have a 
generic Hawking temperature
$T_h$, not necessarily equal to the real temperature $T$. In Euclidean coordinates this leads to a conical
singularity localized at the horizon. In fact, after the substitution $(\rho - \rho_h)/\rho_h = y^2/L^2$, 
by keeping only the leading term in $y$ near the horizon and by neglecting the spatial coordinates in 
(\ref{eq:AdSS}) one gets
\be
\label{eq:horcone}
d s^2 = \frac{4 \rho_h^2 y^2}{L^4} d t^2 + d y^2        \; ,
\end{equation} 
which is the metric of a cone of angle $\alpha$, with $\sin \alpha = \rho_h / \pi T L^2 = T_h/T$.
The integral of the scalar curvature extended to a neighborhood of the horizon can be evaluated
by regularizing the cusp with a spherical cap of small radius $r$, area $2 \pi r^2 (1- \sin \alpha)$ and
constant curvature $2/r^2$. The integral is thus $r$-independent and equals $4 \pi (1- \sin \alpha)$. 
From the gravity action (\ref{eq:5Daction}) we see that
the contribution of the conical singularity to the free energy density (per 3D volume) is
\be
\label{eq:Fcone}
F_{\rm cone} = - T \: 8 \pi M^3 \left(1- \frac{T_h}{T}\right) \frac{\rho_h^3}{L^3} = 
8 \pi^4 \: (M L)^3 \:  T_h^4 \: \left( 1-\frac{T}{T_h} \right)   \; , 
\end{equation}
where we used the fact that the free energy is $-T \cdot S$.

The remaining part of the free energy has to be computed as difference of the gravitational action
(\ref{eq:5Daction}) between the AdS-S solution and pure AdS \cite{Witten:1998zw}. 
Both actions are divergent, but one can cut the $\rho$-integral off at $\rho=\Lambda$, 
take their difference (which is finite) and then let $\Lambda$ go to infinity. In performing this 
limit we can neglect the contribution at the boundary (the surface integral
of the trace of the extrinsic curvature) because it gives a contribution
to the difference $F_{\rm AdS}-F_{\rm AdS-S}$ which goes like $\rho_h^8/\Lambda^4$
and vanishes when $\Lambda\to \infty$. The same result can be obtained regularizing the solutions
with the explicit introduction of the Planck brane. 
Einstein equations force the curvature $R$ to equal $-20/L^2$, so the action reduces to
\be
S  = 2 M^3 \int d^5 x \:\sqrt{-g}( R + 12
k^2)= - \frac{16 M^3}{L^2} \int \sqrt{-g} \: d^5 x \; ,
\end{equation}
thus giving 
\begin{eqnarray}
F_{\rm AdS-S} & = & \frac{16 T M^3}{L^2} \int_0^{1 / T'} dt \int_{\rho_h}^\Lambda d\rho \frac{\rho^3}{L^3} \\
F_{\rm AdS} & = & \frac{16 T M^3}{L^2} \int_0^{1 / T} dt \int_0^\Lambda d\rho \frac{\rho^3}{L^3} 
\end{eqnarray}
as contributions to the free energy densities.
Note that the two time integrals above are extended to different domains: in order to compare the two solutions
one has to impose that the geometry induced on the cutoff surface $\rho=\Lambda$ is the same \cite{Witten:1998zw}.
This implies the following relation:
\be
\label{eq:betas}
\sqrt{\frac{\Lambda^2}{L^2} -\frac{\rho_h^4/L^2}{\Lambda^2}} \frac{1}{T'} = \frac{\Lambda}{L} \frac{1}{T} 
\quad\Rightarrow\quad \left(1- \frac{\rho_h^4}{2 \Lambda^4}\right) \frac{1}{T'} \simeq \frac{1}{T}      \; .
\end{equation} 
The difference between the two free energies is thus given by
\be
\label{eq:Fdifference}
F_{\rm AdS-S} - F_{\rm AdS} = \frac{16 M^3}{L^2} \left[\frac{1}{4 L^3} (\Lambda^4 - \rho_h^4) - \frac{\Lambda^4}{4 L^3}
\left(1- \frac{\rho_h^4}{2 \Lambda^4}\right)\right] = - 2 \pi^4 (M L)^3 T_h^4           \; .
\end{equation}
Adding the singular contribution (\ref{eq:Fcone}) we get the total $F$ for the black hole solution,
\be
\label{eq:totalF}
F = 6 \pi^4 (M L)^3 T_h^4 - 8 \pi^4 (M L)^3 T T_h^3     \; ,
\end{equation} 
where we can distinguish the energetic and the entropic contribution. Obviously 
$F$ is minimum for $T_h=T$: the
Einstein equations forbid the conical singularity in the absence of a source term. The value of $F$ at $T_h=T$ is
\be     \label{eq:Fmin}
F_{\rm min} =  - 2 \pi^4 (M L)^3 T^4            \; .
\end{equation} 
By holography, this equation is interpreted as the free energy
of a strongly coupled large $N$ CFT, with $N^2 = 16 \pi^2 (ML)^3 +1$ \cite{Gubser:2001vj}.

Eq.~(\ref{eq:Fmin}) represents the classical gravity action. However even after including quantum
(or string) corrections the action will keep going like $T^4$. This is because $T$ is not reparametrization invariant.
In particular under the rescaling $(x,t,1/\rho)\to \lambda (x,t,1/\rho)$ we have $T\to T/\lambda$ so 
that the action $\propto \int T^4 d^4x$ is invariant, but any other power of $T$ would not. Then graviton loops
just correct $F$ by factors of $1/(ML)^3\sim 1/N^2\ll 1$.

One can reason similarly for the RS solution with the TeV brane at $\rho=\rho_1=Le^{-kz_1}$.   
 At the classical level one easily finds $F_{\rm RS}-F_{\rm AdS}=0\;$\footnote{This corresponds to the cosmological
constant being tuned to zero and the radion potential  being flat in the RS model.}. Then eq.~(\ref{eq:Fmin})
gives truly $F_{\rm AdS-S}-F_{\rm RS}\simeq -(\pi^2/8) N^2 T^4$ so that at the level
of the classical solutions  AdS-S is thermodynamically favored at
any temperature $T>0$. As we discussed in the previous section, higher order corrections will
give rise to a non trivial $F_{\rm RS}(T,\mu)=T^4 f(T/\mu)$.  Two possibilities 
are then given. In the first, $F_{\rm RS}$
does not even have a stationary point in $\mu$, in which case AdS-S is the only possible phase.
In the second, the stationary value  $\langle\mu\rangle$ is given by the only reasonable scale in the problem,
the scale at which physics becomes Planckian and quantum corrections are unsuppressed $\langle \mu \rangle \sim T/(M L)$. 
In this case the free energy will be very roughly given by the number $T/\mu=M L$ of excited
KK modes: $F_{\rm RS}\sim -(M L)T^4\sim -N^{2/3}T^4$. At large $N$ also 
this second possibility is thermodynamically disfavored with respect 
to AdS-S.

We will now see how radius stabilization changes the picture.



\section{The effect of the Goldberger-Wise field}
The radion modulus of the RS model must be stabilized in order to get a viable solution to the hierarchy problem.
Goldberger and Wise  \cite{Goldberger:1999uk} have shown that this can be done without fine tunings through the
introduction of a bulk scalar field $\phi$. Its action is given by 
\be
S_{\rm GW} = \int d^4 x d \rho \left \{ 
\sqrt{-g} \left[- g^{MN} \pa_M \phi \pa_N \phi - m^2 \phi^2 \right] +
\delta (\rho - \rho_0) \sqrt{-g_0} {\cal{L}} _0 +
 \delta (\rho - \rho_1) \sqrt{-g_1} {\cal{L}} _1
\right \}       \; .
\end{equation}
The brane Lagrangians ${\cal{L}}_{0,1}$ are assumed to force $\phi$ to have the boundary values $\phi(\rho_{0,1})
= L^{-3/2} v_{0,1}$.
The general $\rho$-dependent solution of the equations of motion is
\be     \label{eq:phi}
\phi(\rho) = A \rho^{-4-\epsilon} + B \rho^\epsilon     \; ,
\end{equation}
where $\epsilon = \sqrt{4+m^2 L^2} \simeq m^2 L^2 /4$ for a small mass. $A$ and $B$ are fixed by the 
boundary conditions on the branes.
The induced 4D potential for the radion field $\mu \equiv \rho_1 / L^2$ is \cite{Goldberger:1999uk}
\be     \label{eq:GWpot}
V_{\rm GW}(\mu) = \epsilon v_0 ^2 \mu_0 ^4 +
\left[(4+2\epsilon)\mu^4(v_1-v_0(\mu/\mu_0)^\epsilon)^2-\epsilon v_1^2\mu^4\right]      
+ {\cal{O}} (\mu^8 / \mu_0 ^4)          \; ,
\end{equation}
where $\mu_0 = \rho_0 / L^2$ and we have assumed $| \epsilon | \ll 1$.
For $\epsilon > 0$ the  potential above has a global minimum at
\be     \label{eq:mutev}
\mu = \mu_{\rm TeV} \equiv \mu_0 \left( \frac{v_1}{v_0} \right)^{1/\epsilon} \left[
\frac{8+6 \epsilon + \epsilon^2 + (2+\epsilon) \sqrt{4 \epsilon + \epsilon^2}}{8+8 \epsilon +2\epsilon^2}
\right]^{1/\epsilon}            \; .
\end{equation}
The huge hierarchy between the weak and Planck scale can be naturally obtained
for parameters not far from 1 (e.g. $v_1/v_0 \sim 1/10$ and $\epsilon \sim 1/20$).
For $\epsilon<0$ the only minimum of the potential (\ref{eq:GWpot}) is 
at $\mu=0$. However, with a small change of the TeV
brane tension $\delta T_1$ the potential is modified by a term $\delta T_1 \mu^4$ which can induce a non-trivial 
minimum at $\mu \sim \mu_0 (v_0/v_1)^{1/|\epsilon|}$: a viable solution can be obtained for $v_0<v_1$ 
\cite{Arkani-Hamed:2000ds, Rattazzi:2001hs}.

The reason why this mechanism solves the hierarchy problem is that the potential for the radion field is of
the form $\mu^4 P(\mu^ \epsilon)$, where $P$ is a polynomial: the scale non-invariance is only induced by
the slow varying $\mu^ \epsilon$ term.
The holographic dual of the GW field is an almost marginal operator ${\cal{O}}$ with conformal 
dimension $4+\epsilon$; from this point of view the hierarchy problem is solved through the slow RG 
evolution of ${\cal{O}}$ which dynamically generates a small mass scale $\mu_{\rm TeV}$ \cite{Rattazzi:2001hs}.
This is the same phenomenon of dimensional transmutation at work in technicolor and in the 
Coleman-Weinberg mechanism. A completely general potential of the form $\mu^4 P(\mu^\epsilon)$ can be  
obtained by considering a self interacting GW field \cite{Rattazzi:2001hs};  
for simplicity we stick to (\ref{eq:GWpot}): further terms would not change our conclusions.

The effects of  the introduction of the GW field are calculated neglecting the back-reaction of its
stress-energy tensor on the AdS metric. This is allowed for sufficiently small values of $\phi$
\cite{Goldberger:1999uk}, namely
\be
\label{eq:noback}
v_{0,1} \ll (ML)^{3/2} \sim N   \;.
\end{equation}
In the 4D description this is equivalent to the requirement that the deformation of the CFT induced  
by the operator ${\cal{O}}$ can be treated perturbatively at all scales down to 
 $\mu_{\rm TeV}$.
 
For $\epsilon > 0$ the operator ${\cal{O}}$ is weak in the IR and gets strong in the UV; 
a more conventional situation arises for  
$\epsilon < 0$ in which the deforming operator gets strong in the IR and would eventually completely
spoil  conformal invariance at a scale
\be
\label{eq:Lambda}
\Lambda \simeq \mu_0 \left(\frac{4\pi v_0}{N}\right)^{1/|\epsilon|}
\end{equation}
which is much less than $\mu_{\rm TeV}$ in our weak coupling assumption (\ref{eq:noback}).

At low  enough temperature the free energy in the phase with the TeV brane is well approximated by 
(\ref{eq:GWpot}): it will now surely have a stationary point (a minimum indeed). A phase transition 
will then happen if the  RS-GW free energy becomes lower than that for AdS-S.

Note that eq.~(\ref{eq:GWpot}) has a secondary 
local minimum at $\mu=0$, which describes a full AdS space.  
Full AdS is stable at $T=0$ also with respect to formation of the black hole as
evident form eq.~(\ref{eq:totalF}). Indeed we may consider the case of $\mu=0$ to coincide also with
the $T\to 0$ limit of AdS-S. In the case $\epsilon < 0$ we have seen that we are forced to add 
a contribution $\delta T_1 \mu^4$ to have a minimum in the TeV region: this term must satisfy 
$\delta T_1 < \epsilon v_1^2$ to get the minimum and $\delta T_1 > - (4 + \epsilon) v_1^2$ to have 
a potential 
bounded from below. Even if we cannot make
explicit calculations, we expect that a secondary local minimum is always present in
the small $\mu$ region: the free energy will be modified under the scale
$\Lambda$ by contributions of the order $\Lambda^4$, so that the maximum of
the GW potential at $\mu \lesssim \mu_{\rm TeV}$ will always be high enough to give
a stable minimum near the origin. For $\epsilon > 0$ the presence of a second minimum at $\mu \simeq 0$ 
is not compulsory: adding the term $\delta T_1 \mu^4$ to eq.~(\ref{eq:GWpot}) with $\delta T_1 < - (4+\epsilon) v_1^2$ 
the GW potential remains bounded from below but has a unique minimum.


In order to compare the two free 
energies we must first take the effect of the GW field in the AdS-S background into account . This 
corresponds to evaluating the GW scalar action at the stationary point around the AdS-S background. 
Notice that the action  now contains only the contribution of the Planck brane:
\be     \label{eq:actAdSS}
S_{\rm GW} = \int d^4 x d \rho \left \{ 
\sqrt{-g} \left[- g^{MN} \pa_M \phi \pa_N \phi - m^2 \phi^2 \right] +
\delta (\rho - \rho_0) \sqrt{-g_0} {\cal{L}} _0  \right \}       \; .
\end{equation}
The boundary condition at the TeV brane is replaced with the requirement that the
solution is regular at the Schwarzschild horizon.
Setting $L=1$ the equation of motion for a $\rho$-dependent $\phi$ in the metric (\ref{eq:AdSS}) is
\be
\label{eq:GWeq}
(\rho^5-\rho\rho_h^4)\pa_\rho\pa_\rho\phi+(5\rho^4-\rho_h^4)\pa_\rho\phi-m^2\rho^3\phi=0 \;.
\end{equation}
With the change of variable $z\equiv(\rho/\rho_h)^4$ we obtain the hypergeometric equation
\be
\label{eq:hyp}
z(1-z)\phi'' +(1-2z)\phi'+\frac{m^2}{16}\phi=0   \;.
\end{equation}
The equation is invariant under $z \rightarrow 1-z$ and its solutions are \cite{grad}
\be
\label{eq:hypsol}
F(\alpha,\beta,1,z)\qquad{\rm and}\qquad F(\alpha,\beta,1,1-z)  \; ,     
\end{equation}
with $\alpha$ and $\beta$ solutions of the quadratic equation $x^2-x-m^2/16=0$. 
Hypergeometric functions have a cut on the real
axis from 1 to $+\infty$, so that only the second solution is regular at the horizon; the first one
diverges logarithmically and gives a divergent contribution to the action. 
The regular solution behaves asymptotically ($\rho \gg \rho_h$)  as 
\be
\label{eq:solasy}
\phi (\rho) \simeq A\left [(1+{\cal{O}}(\epsilon^2)) (\rho/\rho_h)^\epsilon - \frac{\epsilon}{8}(\rho/\rho_h)^{-4+\epsilon}
+ \frac{\epsilon}{8}(\rho/\rho_h)^{-4-\epsilon} \right ]\; .
\end{equation}
Note the presence of the second term, which is absent in the solution over 
pure AdS of eq.~(\ref{eq:phi}). 
The overall factor $A$ is fixed by requiring $\phi(\rho_{0})=v_0 / L^{3/2}$, as forced by ${\cal{L}}_0$.
Upon integration by parts,  the GW action (\ref{eq:actAdSS}) is just 
due to boundary terms, as the bulk term vanishes on the equations of motion. 
Moreover, the contribution at the horizon vanishes for the regular solution 
since $g^{\rho\rho}|_{\rho_h} = 0$. Then, after properly 
rescaling the temperature according to  eq.~(\ref{eq:betas}), we find that 
the GW field induces a correction   
\be
\label{eq:Plcont}
\Delta F_{\rm GW}=\sqrt{-g} \;g^{\rho\rho} \pa_\rho\phi \;\phi \Big|_{\rho_0} = \epsilon v_0^2 \mu_0^4 
- \epsilon \frac{\pi^{4+2\epsilon}}{2} v_0^2 T_h^4 \left(\frac{T_h}{\mu_0}\right)^{2\epsilon}
\end{equation}
to the AdS-S action in eq.~(\ref{eq:Fmin}).
The first term can be, as always, canceled by a local counterterm which
redefines the Planck brane tension; the other term depends as expected only on the RG-invariant quantity 
$v_0 \mu_0^{-\epsilon}$. 

\section{Computation of the transition temperature}
We have now all the ingredients to compare the free energies for the two solutions and find in which regime
each one is dominant. 
We refer for simplicity to the original GW case $\epsilon >0$ and with no further modifications of the TeV brane tension;
similar results hold in other cases ($\epsilon <0$).
The contribution $\epsilon v_0^2 \mu_0^4$ due to the boundary term of the GW field on the
Planck brane is common to the two solution and cancels out. When the temperature is sufficiently 
below the weak scale, the free energy in the RS case is well approximated  by the GW
potential (\ref{eq:GWpot}), so that we have to compare
\begin{eqnarray}
F_{\rm RS} & = & \left[(4+2\epsilon)\mu^4(v_1-v_0(\mu/\mu_0)^\epsilon)^2-\epsilon 
v_1^2\mu^4\right]  + {\cal{O}}(T^{4})
                \label{eq:RSpot}                       \\
F_{\rm AdS-S} & = & 6 \pi^4 (M L)^3 T_h^4 - 8 \pi^4 (M L)^3 T T_h^3 
-\epsilon \frac{\pi^{4+2\epsilon}}{2} v_0^2 T_h^4 
\left(\frac{T_h}{\mu_0}\right)^{2\epsilon}   \label{eq:BHpot}   \;.
\end{eqnarray}
The last term in the AdS-S case can be neglected in the weak GW coupling limit of eq.~(\ref{eq:noback}).

The value of the GW potential at the minimum (\ref{eq:mutev}) is
\be 
\label{eq:minimum}
V_{\rm min} \simeq -\epsilon \sqrt{\epsilon} v_1^2 \mu_{\rm TeV}^4 
\end{equation}
which has to be compared with the minimum of $F_{\rm AdS-S}$ (\ref{eq:Fmin}). These quantities will be 
equal at a critical temperature $T_{c}$ given by  
\be
\label{eq:equilibrium}
-\epsilon^{3/2} v_1^2\mu_{\rm TeV}^4 = - 2 \pi^4 (M L)^3 T_c^4 \quad\Rightarrow\quad T_c =  
\left(\frac{8 \epsilon^{3/2} v_1^2}{\pi^2}\right)^{1/4}\frac{1}{\sqrt{N}} \mu_{\rm TeV}   \;.
\end{equation}
Since $\epsilon \sim 1/20$ and since $v_1\ll N$, in order to have negligible back-reaction 
(see (\ref{eq:noback})), we have
\be
\label{eq:limit}
T_c \ll \mu_{\rm TeV}    \; .
\end{equation}
At $T<T_{c}$ the system is in the RS phase. At $T=T_{c}$ the system undergoes a first order phase 
transition to a hot conformal phase. Notice that
by the low value we found for $T_c$   our neglect of thermal corrections to 
eq.~(\ref{eq:RSpot}) was justified. However, when there is a large number $g_{*}$ of   light degrees of freedom
localized on the TeV brane, their contribution to the RS free energy $\sim - g_{*} T^{4}$ may be 
relevant and $T_{c}$ increased. As we will discuss later on $g_{*}$ in the SM is not large
enough to dramatically change the picture. Therefore we will stick to the simple case where the TeV brane 
is empty.

We conclude that if the Universe
was ever heated at temperatures well above the weak scale $\mu_{\rm TeV}$, 
its expansion would be 
driven in that regime by hot CFT radiation.
The transition to the RS solution would only  happen at a temperature below $T_c$. In the case of a 
small back reaction, eq.~(\ref{eq:limit})  implies that the KK gravitons which represent the collective excitations 
of the broken CFT and whose lowest mass is  ${\cal{O}}(\mu_{\rm TeV})$
are never thermally excited. The Universe never experiences a 5D behavior.

\section{Dynamics of the phase transition}
The two gravitational solutions we described are local minima of the free 
energy: we expect a first order phase
transition  proceeding, from a 4-dimensional  perspective, through bubble nucleation. The
bubble will interpolate between the unbroken hot  CFT at infinity and the broken phase inside. From the 
5-dimensional perspective we view this process as the formation of spherical brane patches on the horizon. 
These  expand and  eventually coalesce to form a complete 3-brane.

For the cases in which there is a secondary local minimum at  $\mu=0$, in order to
have a viable cosmology we must require that the rate of bubble nucleation per unit volume $\Gamma$ is larger
than $H^4$ at $T \sim T_c$. If this were  not true, then the Universe would cool down below $T_c$ and 
begin to inflate. This is because the cosmological constant for the false vacuum is positive if  we assume it to be zero in 
the true RS vacuum. Thus  $\Gamma < H^{4}$ corresponds to an old inflation scenario, which is ruled out because of 
its inefficient reheating  \cite{Guth:1983pn,Hawking:1982ga}. In this case of small transition rate the Universe 
would asymptotically approach the local minimum at $\mu = 0$. This
asymptotic cold state is not possible for the cases where there is a unique minimum of the potential: a small
transition rate will cause a period of inflation, but at sufficiently low temperature the Universe will evolve 
towards the unique minimum. We will come back to this possibility in the conclusions. In what follows 
we study the transition rate and the cosmology for the case in which there is a secondary local minimum.

In order to estimate $\Gamma$ it is reasonable to neglect 4D gravity by moving the Planck 
brane to infinity. In the semiclassical approximation to evaluate
 $\Gamma$ one should normally find the 5D gravity solution corresponding to 
 the bounce. We will not undertake this task and not just because it is 
 technically hard. The main reason is that the exact bounce solution  would 
 necessarily lead us out of 
 the limited domain where the effective field theory description of the 
 RS model applies.  As it will be clarified below, by topology  the bounce will
 involve regions of space where the Euclidean time cycle is of
 (sub)Planckian length: here quantum gravity effects are unsuppressed 
 and we also expect the unknown physics that resolves the brane to 
 matter. However we will argue that when the critical temperature is 
 low enough, the dominant contribution to the tunneling amplitude will 
 come from the region $\mu \gg T$ where effective field theory applies 
 and where the radion is the only relevant mode.
 In this regime we will be able to give a reasonable estimate of 
 $\Gamma$.

\begin{figure}[t]             
\begin{center}
\includegraphics[width=10cm]{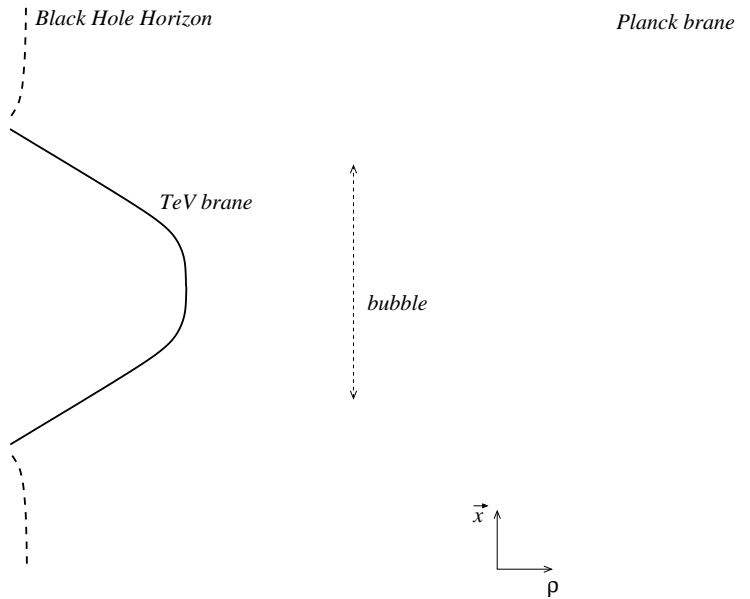}
\caption{\label{fig:bolla} 5D picture of the 4D bubble configuration. The two solutions are connected by
moving the black hole horizon and the TeV brane towards the AdS infinity.}
\end{center}
\end{figure}

\begin{figure}[t]             
\begin{center}
\includegraphics[width=14cm]{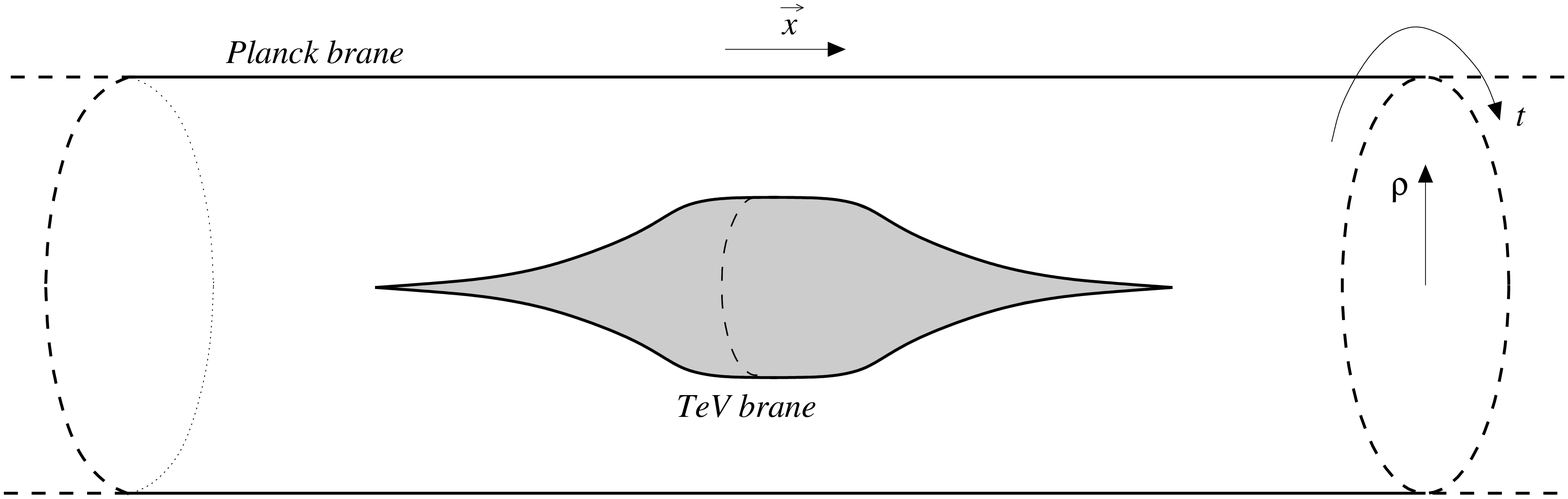}
\caption{\label{fig:topologia} The topology of the bubble. We have suppressed two spatial dimensions, so that
the AdS-S solution is a cylinder: the bubble appears as hole in the cylinder. The rotational invariance is here seen 
as a discrete ${\mathbb Z}_2$ inversion symmetry.}
\end{center}
\end{figure}

Let us now consider the general features of the bounce.
Note, first of all, that the two gravitational solutions we want to interpolate between have different 
topologies. The AdS-Schwarzschild is equivalent to the product of a disk ($\rho$ is the radial variable and $t$ is 
the angular one) bordered by the Planck brane and ${\mathbb R}^3$ (the three spatial dimensions); to obtain 
the RS space we have to insert the TeV brane: we cut a hole in the $\rho$ - $t$ disk making an annulus. 
So AdS-S is simply connected while RS is not.
It is well known that in General Relativity a topology change can be accomplished without generating
a configuration of infinite action. In fact the two solutions
can be deformed one into each other by moving the horizon towards $\rho = 0$ and then moving the brane back 
from $\rho = 0$. In this deformation we pass through ordinary AdS space with periodic time. This has the same topology 
as AdS-S except that all the points at $\rho = 0$ are identified. All 
the configurations along the path have finite action per unit volume. 
Therefore we  find it plausible to assume that there will exist a bubble  solution (instanton)
performing such an interpolation:
in moving from the outside to the center of the bubble we see the horizon going towards $\rho = 0$, 
we arrive at the pure AdS and then the 
brane comes in from $\rho = 0$ to a finite value, as shown in fig.~\ref{fig:bolla}. In figure \ref{fig:topologia} 
we show what the bubble looks like topologically.     

In other words, we have chosen a particular one-parameter set of configurations: the RS model with generic
TeV brane position $\mu$ is glued to the set of black hole configurations, parametrized by their Hawking 
temperature $T_h$, through the pure AdS solution, which is obtained both for $\mu \to 0$ and for $T_h \to 0$. 

\begin{figure}[t]             
\begin{center}
\includegraphics[width=14cm]{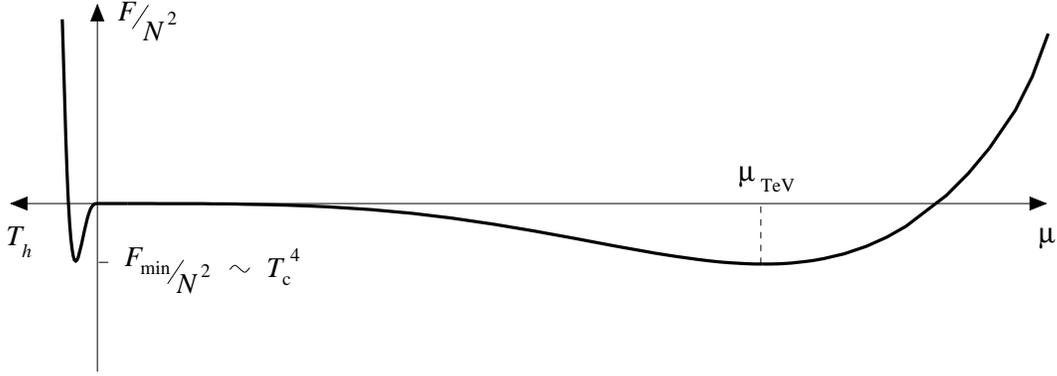}
\caption{\label{fig:gluedpot} The free energy for the RS solution (right side) and the black hole one (left).
RS model is parametrized by the radion VEV $\mu$, while the AdS-S one by the Hawking temperature $T_h$.
The two classes of solutions coincide at $\mu = T_h = 0$.
}
\end{center}
\end{figure}

A potential for this set of configurations is obtained by joining (\ref{eq:RSpot}) and (\ref{eq:BHpot}) 
 (see figure \ref{fig:gluedpot})  
\begin{eqnarray}
F_{\rm RS} & = & \left[(4+2\epsilon)\mu^4(v_1-v_0(\mu/\mu_0)^\epsilon)^2-\epsilon v_1^2\mu^4\right] \quad {\rm and} 
\label{eq:Vbrane}\\
F_{\rm AdS-S} & = & 6 \pi^4 (M L)^3 T_h^4 - 8 \pi^4 (M L)^3 T T_h^3 
\end{eqnarray}
(we neglect the GW contribution in the black hole case, as it is subleading in $N$). 
This potential 
should be taken with some caution. Indeed in  
the region $\mu < T / ML$ quantum gravity  corrections are unsuppressed and also our effective 
field theory description of the brane breaks down. In this region we  expect  the physics that 
resolves the brane to become important. On the black hole side it is also not obvious that the relevant 
degree of freedom is simply represented by the horizon position. On the other hand in the region $\mu > T/ ML$ the radion (and its potential) gives a good description of the smooth brane deformations, 
for which $d\mu/dx\ll \mu^{2}$. When  we can treat the brane by effective field theory
the radion is also the lightest mode:  unlike the KK masses, the GW potential is suppressed by $v_{1}^{2}/N^2 \ll 1$.
Now, our crucial remark is that in the tractable case where $T_{c}\ll \mu_{\rm TeV}$, the 
parametrically dominant  contribution to the bubble action  comes from the region of large $\mu$, 
with small $d(1/\mu)/dx$,
where we can control our effective action. The modeling of the potential in the region
of small $\mu$ and on the horizon side will not sizably affect the result. The basic reason for 
this is gotten by looking at fig.~3. The normalized potential is very flat on the brane side, where 
the depth is only of order $-(v_{1}/N)^{2}\mu_{\rm TeV}^{4}«$ over a ``large'' distance $\sim \mu_{\rm TeV}$.
On the black hole side there is no small parameter suppressing the depth. Then in order to balance 
the gradient energy due to the large distance in $\mu$ the bubble will have to be large and therefore 
most of its action will come from the large $\mu$ region. This will become more clear in the 
quantitative discussion below. Also fig.~\ref{fig:bolla} is useful to  get an idea: 
as $T_{c}\ll \mu_{\rm TeV}$, the brane patch should protrude well out into the bulk, thus dominating the 
action.

A further point has to be stressed in the case $\epsilon <0$. In this case the operator which induces the breaking 
of the conformal symmetry gets strong in the IR and at the scale $\Lambda$ defined in (\ref{eq:Lambda}) the
conformal picture is completely spoiled: correspondingly in the gravity picture the GW field leads to a strong 
deformation of  the pure AdS metric. This implies that in both phases we expect to enter an unknown regime 
at $T_{h} < \Lambda$ and $\mu < \Lambda$, where the potential will be of size $\Lambda^{4}$ instead 
of naively going through zero. However, when $\Lambda\ll T_{c}$, for which we can perturbatively 
study the phase transition, the region of field space affected by large non perturbative corrections 
is small and  makes only a small correction to the bounce action, which as we said is dominated 
by the large $\mu$ region.
In conclusion we expect that, for a sufficiently weak deforming operator ${\cal{O}}$, the transition has the same 
characteristics as for $\epsilon > 0$. 

We now move to estimate the transition rate. The bubble nucleation rate per
unit volume in a first order phase transition can be written 
\be
\label{eq:bubblerate}
\Gamma = \Gamma_0 e^{-S}   \;;
\end{equation}
$\Gamma_0$ is a determinant prefactor which in our case we expect to be of the order TeV$^4$. 
The dominant dependence on the model parameters 
is encoded in the exponential of the Euclidean action $S$, computed on the bubble solution 
interpolating between the false and the true minimum. If the temperature is high enough the characteristic length
scale of the bubble is much larger than the time radius $1/T$, so that the favored instanton will have an $O(3)$
symmetry and $S$ reduces to $S_3/T$, where $S_3$ is the spatial Euclidean action. At low temperature an $O(4)$
symmetric instanton will instead be dominant.     

A subtle point should be stressed: in order to compute the exact bounce solution one should be able to evaluate 
the action for a configuration with a space dependent $T_h (\vec x)$. To do that one should find a complete solution 
of the Einstein equations describing the bubble. As this is an almost impossible task we must face the fact that we 
do not know the ``kinetic'' term
of the $T_h$ field. Note that this implies that we are not able to ``canonically'' normalize the horizontal scale 
in the left side of figure \ref{fig:gluedpot}. For the other side of the figure the situation is more conventional, as
we know that the radion kinetic   term  is
\cite{Csaki:2000mp,Goldberger:2000un,Rattazzi:2001hs}
\be
\label{eq:radionk}
{\cal{L}_{\rm kin}} = -12 \sqrt{-g} (ML)^3 (\pa\mu)^2  \;.
\end{equation}
The hierarchy $\mu_{\rm TeV} \gg T_c$ between 
the two characteristic scales in the potential allows to establish the main features of the bounce 
action $S$.

Even if we do not know the explicit form of the contribution to the action $S$ from the ``black hole'' side of the 
instanton, we know that this is purely gravitational, so that it will be proportional to $(ML)^3 \sim N^2$ and it
will be characterized by the unique energy scale $T \sim T_c$. 
The kinetic term of the radion $\mu$ in (\ref{eq:radionk}) is similarly proportional to $(ML)^3 \sim N^2$, 
so we can factor $N^2$ out of  the entire action by rescaling the stabilizing GW potential 
(\ref{eq:Vbrane}): $F_{\rm RS} \equiv N^2  \tilde F_{\rm RS}$, $F_{\rm AdS-S} \equiv N^2  \tilde F_{\rm AdS-S}$.
The rescaled potential $\tilde F_{\rm RS}$ has a characteristic horizontal scale 
$\sim \mu_{\rm TeV} \gg T_c$ and depth $\sim T_c^4$, while, as we said, $\tilde F_{\rm AdS-S}$ 
has $T_c$ as the only typical scale.  
At leading order in $N$ we will estimate the action $S$ neglecting the width of $\tilde F_{\rm AdS-S}$ with respect
to the much greater $\tilde F_{\rm RS}$ one: the different horizontal scales characterizing the two parts of the 
potential allows us to disregard, at leading order in $N$, the unknown ``kinetic'' term for the black hole horizon.

Within the above approximation, it is useful to rescale the radion field 
\be
\label{eq:mutilde}
\tilde \mu \equiv \mu \;\epsilon^{3/8}\sqrt{\frac{v_1}{N}}  \; .
\end{equation}
In terms of $\tilde \mu$, the width of the potential $\tilde\mu_{\rm TeV}=\mu_{\rm TeV}\epsilon^{3/8}\sqrt{{v_1}/{N}}$
also controls its depth $\tilde F_{\rm RS \; min} \simeq -\tilde \mu_{\rm TeV}^4 $.
In this way the position of the minimum $\mu =\mu_{\rm TeV}$ is rescaled to the same energy scale of the 
potential depth $\tilde F_{\rm RS \; min} \simeq -\epsilon \sqrt{\epsilon} v_1^2 \mu_{\rm TeV}^4 /N^2$. 
To canonically normalize the kinetic term for $\tilde \mu$ (see (\ref{eq:radionk}))
we have to redefine the coordinates: $\tilde x \equiv x \cdot 2 \pi/\sqrt{3} \cdot \epsilon^{3/8} \sqrt{v_1/N}$. 
In such a way we have factored out all the relevant parameters from the action $S$. In the limit of small and 
high temperature we expect
\be \label{eq:actions}
S_4 \sim \frac{N^4 3^2}{(2 \pi)^4 \epsilon^{3/2} v_1 ^2} \tilde S_4 \left( \frac{T}{T_c} \right)
\qquad\qquad
\frac{S_3}{T} \sim \frac{N^{7/2} 3^{3/2}}{(2 \pi)^{3} \epsilon^{9/8} v_1^{3/2}} \tilde S_3 
\left( \frac{T}{T_c} \right) \;,
\end{equation}
where the functions $\tilde S_{3,4} (T/T_c)$ have no small parameters. The characteristic length scale
of the bubble in the original coordinates $x$ is $R_{\rm bubble} \sim \mu_{\rm TeV}^{-1} N / (v_1 \epsilon^{3/4})$: 
the variation of the brane position is slow and the condition $d \mu/ d x \ll \mu^2$ is satisfied. 
For $T \gg R_{\rm bubble}^{-1} \sim T_c \sqrt{v_1} \epsilon^{3/8} / \sqrt{N}$ we expect the thermal
bounce to be favored with respect to the $O(4)$ symmetric solution.

The estimates (\ref{eq:actions}) can also be seen in a rather different 
way. Essentially they are obtained by neglecting
all what happens in the unknown region $\mu < T /ML$, as if the black hole minimum were located just at the end 
of the region under the control of effective field theory. The important point is that the addition of the
unknown region of potential can only make the transition slower. 
This conclusion is obvious if the problem of bubble nucleation is formulated in terms of tunneling through a 
barrier of potential and gradient energy \cite{coleman}: our approximation corresponds to setting  the
gradient energy to zero and the potential to a constant $V_{0}\equiv 
F_{\rm min}$ in the unknown region.
Therefore our estimates give both the transition rate at leading order in $N$ and, disregarding the unknown region 
of potential, an upper limit on the transition rate itself. This makes more robust the constraints we will derive 
in the next section requiring the transition to be fast enough to avoid old inflation.

The above results, though qualitative, are quite reasonable.
First of all we see that the phase transition gets more and more difficult the
larger $N$: this could be expected as at large $N$ the transition implies a dramatic decrease in the 
number of degrees of freedom.
Also the $v_1$-dependence is not unexpected: $v_1$ describes the strength of the stabilization mechanism, or in the
4D language the influence of the deformation operator introduced in the CFT. As only in presence of such a 
mechanism the transition is possible, we expect that a large $v_1$ makes the transition faster and faster.

In the following section we show that for the rate of bubble nucleation to  be 
cosmologically acceptable 
a small $N$ and a strong stabilization mechanism are required. 

\section{Thin wall approximation}
Let us focus first on a temperature $T$ very close to $T_{c}$ so that the two minima
are almost degenerate and the bubble is described by the thin wall approximation \cite{coleman}.
In this limit we can easily estimate the action for a thermal bounce:
\be
\label{eq:thinwall}
S_3 = \frac{2 \pi}{3} \frac{\left[\int^{\rm \mu_{\rm TeV}}_{\rm 0} d \mu \sqrt{2 
F_{\rm RS}(\mu)}\right]^3}{(\Delta F)^2} \times \left( \frac{3 N^2}{4 \pi^2} \right)^{3/2} \; ,
\end{equation}
where $F_{\rm RS}$ is shifted to be zero at the true minimum.
In the approximation described in the previous section we have neglected the horizontal width of the black
hole side potential thus extending the barrier penetration integral only over the RS side. $\Delta F$ is the free energy difference
between the two minima and the last factor takes into account the non canonical normalization of $\mu$ 
(see (\ref{eq:radionk})). 

We can estimate the various quantities appearing in (\ref{eq:thinwall})  
\begin{eqnarray}
& & \int^{\mu_{\rm TeV}}_{0} d\mu \sqrt{2 F_{\rm RS}(\mu)}  \simeq  \sqrt{2} \epsilon^{3/4} v_1 \mu_{\rm TeV}^3 \\
& & \Delta F = \epsilon^{3/2} v_1^2 \mu_{\rm TeV}^4 \left[1-\left( \frac{T}{T_c} \right)^4\right] \\ 
&  & \frac{S_3}{T}  \simeq  \frac{2 \pi}{3} 2^{3/2} \left( \frac{\pi^2}{8} \right)^{1/4}  
\frac{N^{7/2} 3^{3/2}}{(2 \pi)^{3} \epsilon^{9/8} v_1^{3/2}} 
\frac{ (T_c/T) }{\left[1-\left( T/T_c \right)^4\right]^2}  \;.  
\end{eqnarray}
Evaluating the numerical factors we get
\be
\label{eq:S3final}
\frac{S_3}{T} \simeq 0.13 \times \frac{N^{7/2}}{\epsilon^{9/8} v_1^{3/2}} \times 
\frac{(T_c/T) }{\left[1-\left( T/T_c \right)^4\right]^2}   \;,
\end{equation}
which shows the correct dependence on $N$, $\epsilon$ and $v_1$ found in the previous section.
We have checked by numerical computations that the thin wall approximation gives a reasonable estimate 
(modulo a factor $\sim$ 2) of the action until the energy difference of the two minima is of the same order of 
the barrier height. Therefore the thin wall approximation is suitable for a qualitative discussion.

Taking the (\ref{eq:S3final}) as an estimate of the leading contribution in $N$ and $1/v_{1}$, we can infer
a cosmological constraint on the parameters of the model. 
To avoid the old inflation scenario the transition must take place at $T \sim T_c$, so that the thermal
bounce is favored and the thin wall approximation is reasonable. Requiring $\Gamma > H^4_{T \sim T_c}$
we have
\be
\label{eq:limit2}
0.13 \frac{N^{7/2}}{\epsilon^{9/8} v_1^{3/2}} \lesssim 137 \quad\Rightarrow\quad 
\frac{N}{\epsilon^{9/28} v_1^{3/7}} \lesssim 7 \;,
\end{equation}
where we used  the fact that $(T_c/T) / \left[1-(T/T_c)^4 \right]^2 > 1$ for $T < T_c$ 
(\footnote{We have neglected the logarithmic corrections to (\ref{eq:limit2}) one would obtain considering that
the number of relativistic degrees of freedom $g_*$ is of order $N^2$.}). 
Taking $\epsilon \simeq 1/20$ we finally get 
\be
\label{eq:theend}
\frac{N}{v_1^{3/7}} \lesssim 3   \;.
\end{equation}

In the above computations we used the specific GW potential with $\epsilon > 0$; other possible solutions can be 
obtained by modifying the TeV brane tension and/or taking $\epsilon < 0$. We expect that also in these cases
the constraint (\ref{eq:theend}) will not be significantly modified. For instance, allowing for an 
extra contribution   $\Delta V \sim - \delta T_1  \mu^{4}$ due to the TeV brane tension one has generically
a deeper minimum $V_{\rm min} \sim -\epsilon \delta T_1  «\mu_{\rm TeV}^{4}«$. Then the bound above becomes
\be
\label{eq:ps}
\frac{N}{\delta T_1^{3/14}} \lesssim 3.7   \;.
\end{equation}

The bound (\ref{eq:theend}) shows that a model with a large $N$, which in the 5D picture allows to neglect 
quantum gravity effects in the
AdS solution, with a perturbative GW stabilization mechanism ($v_1 \ll N$) seems to have an unviable high $T$ cosmology.
The introduction of a stronger stabilization mechanism would partially 
alleviate the problem, but one would loose the  nearly-AdS (or nearly conformal) picture.    

Before going to the conclusions, an important comment is necessary. In 
all the calculations we have done so far
we have neglected the contribution to $F$ from the light degrees of freedom confined to the brane. 
In other words we have neglected the Standard Model! The contribution of the brane excitations to $F_{\rm RS}$ 
(eq.~(\ref{eq:RSpot})) is given by 
\be
\label{eq:FSM}
F_{\rm SM} = -\frac{\pi^2}{90} g_{*\; \rm brane} T^4  \;,
\end{equation}  
where $g_* = 106.75$ for the Standard Model particles at high temperatures. 
For the calculation of the transition temperature it is easy to see that eq.~(\ref{eq:equilibrium}) 
holds with the substitution 
\be
\label{eq:substit}
N^2 \rightarrow N^2 -\frac{4}{45} \; g_{*\; \rm brane}   \;:
\end{equation}
there is a smaller difference in the number of degrees of freedom 
between the two phases and $T_c$ gets larger. The inclusion of these light brane excitations gives a 
negligible correction to the transition temperature and to the dynamics of the transition if 
$g_{*\; \rm brane} \ll 45/4 \; N^2$, which holds, in the case of the SM particles, with $N$ much larger than
3. We consider this bound to be rather weak, as for smaller $N$ the gravitational description cannot be trusted.
It is interesting to note that an independent limit to the number of light degrees of freedom confined to the brane
could derive from some ideas on the irreversibility of the RG flow, which, roughly speaking, indicates that the number
of degrees of freedom must decrease going from the UV to the IR \cite{Appelquist:1999hr}.  
If $g_{*\; \rm brane}$ is large enough the 4D theory we are holographically describing has a number of IR degrees 
of freedom larger than in the UV. These bounds could constrain the possibility 
to give a stringy realization of the RS 
model with a large number of light particles confined on the TeV brane.       

\section{Conclusions and outlook}
We have used the holographic interpretation of the Randall-Sundrum model to study its finite
temperature properties. While we think  this problem is interesting {\it per se}, our main motivation
is to understand the early cosmology of the model, when the temperature of the universe was of the order 
or bigger than the electroweak scale. The
picture that emerges is that the model is in different phases at low and high temperature, with a
first order phase transition at a temperature $T_{c}$. At low temperature we have the RS
phase, where physics is described by the SM particles and possibly the lightest KK modes. Here the
free energy is dominated by the dynamics that stabilizes the radion. In
our paper we focused on a minimal GW mechanism, but notice that in general the
SM Higgs  has a part in this dynamics \cite{Rattazzi:2001hs}. So we may more broadly say that in the low temperature phase
radion stabilization and electroweak breaking dominate the free energy.
In the high temperature phase, the breaking of conformal symmetry by both the presence of the TeV
brane (spontaneous) and the presence of the GW coupling (explicit) is screened by thermal effects.
The system behaves here like a hot CFT. In the 5d description the TeV
brane is replaced with the black hole horizon of AdS-Schwarzschild.  Using the gravitational picture
we have calculated the critical temperature $T_{c}$.  We find that when GW mechanism is
associate to a weak coupling (small distortion of AdS metric) $T_{c}$ is parametrically suppressed
with respect to the scale $\mu_{\rm TeV}$ that controls the KK splittings. 
So $T_{c}$ is not bigger than   the Fermi scale, and more likely even somewhat smaller. The phase transition to the low
energy RS regime and electroweak breaking happen essentially at once: at $T<T_{c}$ the gravity
solution with stabilized radion and non zero Higgs VEV becomes thermodynamically favored over AdS-S.

We have then studied how the Universe transits to the SM phase as it cools down during expansion.
The phase transition is first order, so it proceeds through the nucleation of bubbles, their
expansion and final collision. In order to see if this process is fast enough we
have estimated the rate of bubble nucleation in a Hubble volume $\Gamma / H^3$ and compared it to the rate of expansion
$H$. Finding the gravitational bounce solution that controls $\Gamma$ in the
semiclassical approximation is however not just hard but also beyond the limitations
of our effective field theory approach. Indeed, the bounce
would, by topology, involve regions of space where the Euclidean time cycle shrinks to Planckian length:
here effective field theory breaks down, and, among other things, the physics that resolves the TeV brane
becomes surely important. However for the case of a weak GW coupling $v_{1}\ll (M L)^{3/2}$ the
bounce action, which is large, is dominated by the region of field space over which the relevant
gravitational mode is simply the radion. The basic reason for this is that for $v_{1}\ll (M L)^{3/2}$ the
critical temperature is smaller that the KK mass scale $\mu_{\rm TeV}$.
Then there is a big range in field space $T<T_{c}\ll \mu<\mu_{\rm TeV}$ where
the radion is the lightest degree of freedom. The ``brane bubble'' depicted
in Fig.~1 lies mostly in this range. We find that the 3-dimensional bounce action $S_{3}/T$ goes roughly like
$(ML)^{3}\times [(ML)^{3/2}/v_{1}]^{3/2}$ where the first factor comes just from the gravitational
action being $\propto M^{3}$ while the second enhancement is due to the radion dominance we just
mentioned. If the GW mechanism were associated to  a sizeable deformation of the AdS metric around
the TeV brane, {\it i.e.} if $v_{1}\sim (M L)^{3/2}$, then there would be just one
energy scale $T_{c}\sim \mu_{\rm TeV}$. In this case the whole problem should be reconsidered.
In fact, now the contribution of the GW field to the action of the black hole solution
is important and may well make it classically unstable at low enough temperature. Therefore
the transition could now even be of second order. Anyway, even if the transition remained
 of first order, a full gravitational bounce solution,
not just the one for the radion mode, would be needed.  

Given a perturbative stabilization mechanism, if the Big Bang temperature exceeds the Fermi
scale, then the transition to the present cosmological era poses a big constraint on the RS model
parameters. In practice the models in which there is a secondary minimum at $\mu=0$
are forced to be on the verge of being perturbatively untractable:
the GW field must cause a sizeable backreaction on the metric and, more importantly, 
$ML$ has to be so small that we expect no more than a handful
of weakly coupled KK modes. Above these lowest modes  the mode width is so large that they overlap
to form  a continuum. 

However an interesting scenario is possible if we choose the parameters of the
potential in such a way that $\mu = 0$ is not a stable minimum; this is possible when
 $\epsilon>0$ by suitably correcting the TeV 
brane tension with respect to the basic RS value. In this case the small transition rate causes
the Universe to inflate, but at sufficiently low temperature it will be destabilized towards the unique minimum.
We expect that this will happen for $T \lesssim H$: in this regime the gravitational corrections to the transition
rate are important \cite{Coleman:1980aw} and we expect the fluctuations induced by the inflating background to 
be enough 
to overcome the potential barrier. Is this a viable scenario of inflation? If the expansion lasts until $T \lesssim H$ 
we have enough inflation to solve the usual smoothness and flatness problems, but it is hard to understand if the 
model gives a 
phenomenologically acceptable spectrum of density perturbations. If the theory were exactly conformal
we would not have any perturbations, because the inflating Universe has a conformally flat metric. However, 
in our case conformal
invariance is explicitly broken by the slowly running GW coupling: even if this non-conformal 
deformation becomes small in the IR, it runs sufficiently slowly that we do not expect a huge suppression of
the induced perturbations. Further studies are required to understand the viability of this scenario.

The last ``trivial'' possibility is that the Big Bang started out with
temperature at or below the Fermi scale, in which case one can consistently assume to be from the
beginning in the RS minimum of the free energy. This poses no constraints on the parameters, though
our theoretical control of early cosmology is not dramatically different than in other models with TeV scale 
extra dimensions.

The holographic point of view could also be useful to study non-standard cosmological solutions 
with non-thermal, high energy initial conditions. The solution described in the Appendix may be a starting point
for further studies in this direction. 

\paragraph{Acknowledgments} We thank Riccardo Barbieri, Roberto Contino, Luigi Pilo, Alessandro Strumia and 
Alberto Zaffaroni for useful comments.

\begin{appendix}
\section{A radion driven cosmology \label{radiondriven}}
A general solution of the 5D Einstein equations in ${\rm AdS}_5$ in presence of 4D branes can be easily
obtained studying the brane motion in the static background metric \cite{Kraus:1999it}. 
The dynamics of the brane is fixed by Israel's junction conditions relating the discontinuity in the brane's 
extrinsic curvature to its energy-momentum tensor. For the RS model we have two branes moving in the bulk and
we suppose their tension to be fixed to the values $\sigma_{\rm Planck} = - \sigma_{\rm TeV} = 24 M^3 / L$ in
such a way that a static solution in possible. 

It is rather interesting to see that, in absence of a stabilization mechanism, the trivial time independent 
solution is not the only possible. The generic bulk geometry is of the form (\ref{eq:AdSS}) with the presence of
an event horizon at $\rho = \rho_h$. We consider the geometry in which both branes are outside the horizon at
$\rho > \rho_h$ (\footnote{A brane which is inside the black hole horizon is difficult to interpret from the
holographic point of view as the dual theory lives outside the horizon and do not feel what is going on inside.}).
As the branes are situated at the orbifold fixed points, the horizon lies outside the physical space: the only 
difference with the trivial static solution is that the metric between the branes is different, 
{\em as if} a black hole horizon were present behind the TeV brane.

Using Israel's junction conditions, we can obtain the position of each brane as a function of its own proper time
$\rho_{\rm TeV}(\tau_{\rm TeV})$ and $\rho_{\rm Pl}(\tau_{\rm Pl})$. The two branes will obey the same equation
\be
\label{eq:branepos}
\dot\rho^2 - \frac{\rho_h^4/L^2}{\rho^2} = 0 \;,
\end{equation}
where $\rho$ stands either for $\rho_{\rm TeV}$ or for $\rho_{\rm Pl}$ and the derivative is taken with respect 
to the brane proper time. The metric induced on the brane will have the form
\be
\label{eq:induced}
ds^2_{\rm brane} = -d \tau^2 + \rho^2_{\rm brane} (\tau) dx^i dx^i   \;,
\end{equation}
so that eq.~(\ref{eq:branepos}) describes the evolution of the scale factor with time. Properly fixing the origin
of time, the solution of the equation of motion for the Planck brane at a sufficient distance from the horizon is
\be
\label{eq:Plmov}
\rho_{\rm Pl} \simeq \sqrt{\frac{2}{L}} \rho_h \tau_{\rm Pl}^{1/2} \;.
\end{equation}
The evolution of the Planck brane is the same that in absence of the TeV one and describes, in the holographic 
counterpart, a Universe evolving as it were radiation dominated. 

At first, this behavior in presence of two branes may be puzzling. We have a ``radiation dominated'' Universe, but,
as there is no real event horizon, the state has no entropy and is non thermal. To understand what is going on,
we will study the radion evolution in the 5D picture and see that its time dependence gives, in the 4D viewpoint,
the energy density and pressure which drive the Universe RD expansion. 

Even if the two branes follow the same equation of motion (\ref{eq:branepos}),
the radion VEV, which in the parameterization we use in the following is proportional to 
$\rho_{\rm TeV}/\rho_{\rm Pl}$ \cite{Rattazzi:2001hs}, is not constant in time. Let us calculate 
\be
\label{eq:radevol}
\frac{d}{d \tau_{\rm Pl}} \frac{\rho_{\rm TeV}}{\rho_{\rm Pl}} = \frac{d \rho_{\rm TeV}}{d \tau_{\rm Pl}} 
\frac{1}{\rho_{\rm Pl}} - \frac{\rho_{\rm TeV}}{\rho_{\rm Pl}^2} \frac{d \rho_{\rm Pl}}{d \tau_{\rm Pl}} =
\frac{\rho_h^2}{L} \frac{1}{\rho_{\rm Pl}^2} \left(1 -\frac{\rho_{\rm TeV}}{\rho_{\rm Pl}} \right)\;, 
\end{equation}
where we have used the relation $d \tau_{\rm Pl}/\rho_{\rm Pl} \simeq d \tau_{\rm TeV}/\rho_{\rm TeV}$, valid 
for $\rho_{\rm TeV} \gg \rho_h$.
For $\rho_{\rm TeV} \ll \rho_{\rm Pl} $, which in the 4D language means that the radion VEV is much smaller 
than the Planck scale, equation (\ref{eq:radevol}) reduces to
\be
\label{eq:radevol2}
\frac{d}{d \tau_{\rm Pl}} \frac{\rho_{\rm TeV}}{\rho_{\rm Pl}} = \frac{\rho_h^2}{L} \frac{1}{\rho_{\rm Pl}^2}  
= \frac{1}{2 \tau_{\rm Pl}}     \;,
\end{equation}
which shows how the radion VEV evolve with the 4D scale factor and with time.

Now we switch to the 4D picture and check that this behavior can be reproduced by the 4D Einstein equations.
The radion $\phi$ is conformally coupled to gravity ($\xi = 1/6$), its Lagrangian and stress-energy tensor
being \cite{Csaki:2000mp,Goldberger:2000un,Rattazzi:2001hs}
\begin{eqnarray}
{\cal{L}} & = & \sqrt{-g} \left[
2 (ML)^3 \mu_0 ^2 R(g) - \frac{1}{12} \phi ^2 R(g) - \frac{1}{2} (\pa \phi)^2
\right]         \;,     \\
T_{\mu \nu} & = & \pa _\mu \phi \pa _\nu \phi -\frac{1}{2} \eta _{\mu \nu} (\pa \phi)^2
+ \frac{1}{6} (\eta_{\mu \nu} \Box - \pa _\mu \pa _\nu) \phi^2  \; .
\end{eqnarray}
The Friedmann equations for the Hubble parameter $H \equiv \dot \rho / \rho$ 
in the case of a time dependent radion are
\begin{eqnarray}
H ^2 & = & \frac{8 \pi}{3} G_4 \cdot \frac{1}{2} \dot \phi ^2           \label{eq:Fr1}          \\
\dot H & = & - 4 \pi G_4 \left( \frac{2}{3} \dot \phi ^2 - \frac{1}{3} \phi \ddot \phi
\right)         \label{eq:Fr2}          \; ,
\end{eqnarray}
where the 4D Newton's constant is defined by $1/(16 \pi G_4) = 2 \mu_0 ^2 (ML)^3$.
Using the equation of motion for $\phi$,
\be
\ddot \phi + 3 H \dot \phi + \frac{1}{6} R(g) \phi = 0  \; ,
\end{equation} 
we see that the second term in the RHS of (\ref{eq:Fr2}) can be neglected in the limit $\phi \ll M_{\rm Pl}$, already
used in the 5D picture.
Solving for $H$ we obtain
\be
\dot H = - \frac{8 \pi}{3} G_4 \dot \phi ^2 = -2 H^2    \; ;
\end{equation} 
the scale factor evolves as in a RD Universe, 
\be
\rho (\tau) \propto \tau ^{1/2}         \; ,
\end{equation}
consistently with (\ref{eq:Plmov}).
The ratio $\rho _{\rm TeV} / \rho _{\rm Pl}$ can be written in the 4D variables as $\phi \sqrt{4 \pi G_4 / 3}$; its
evolution is given by (\ref{eq:Fr1}),
\be
\sqrt{\frac{4 \pi G_4}{ 3}} \dot \phi = H = \frac{1}{2 \tau}    \; ,
\end{equation}
which is exactly the same result obtained in the 5D picture (\ref{eq:radevol2}). In the above discussion we have
implicitly assumed a radion increasing with time, but a completely equivalent cosmological expansion can be obtained
also with a decreasing radion, which corresponds to the other root of (\ref{eq:branepos}).  

We have checked that the general 5D solution consisting of a slice of AdS-S between the two branes,
in absence of a stabilization mechanism, can be interpreted in the 4D picture as a radion driven cosmology. 
The radion field, being conformally coupled to gravity, has a radiation-like behavior as source of gravity 
thus giving a RD-like cosmological evolution.  

\end{appendix}

\end{document}